# Non-equilibrium heat capacity of polytetrafluoroethylene at room temperature

**Authors: J.-L. Garden, J. Richard, H. Guillou and O. Bourgeois**

*Institut Néel, CNRS et Université Joseph Fourier, BP 166, 38042 Grenoble Cedex 9, France.*

**Abstract**

*Polytetrafluoroethylene can be considered as a model for calorimetric studies of complex systems with thermodynamics transitions at ambient temperature. This polymer exhibits two phase transitions of different nature at 292 K and 303 K. We show that sensitive ac-calorimetry measurements allow us to study the thermodynamic behaviour of polytetrafluoroethylene when it is brought out of thermodynamic equilibrium. Thanks to the thermal modelisation of our calorimetric device, the frequency dependent complex heat capacity of this polymer is extracted. The temperature and frequency variations of the real and imaginary parts of the complex heat capacity are obtained when polytetrafluoroethylene undergoes its first-order structural phase transition at 292 K.*

Keywords: ac-calorimetry; Non-equilibrium thermodynamics; Polytetrafluoroethylene; Polymer; Frequency dependent complex heat capacity; Relaxation time constants.

\* Corresponding author. Fax: 33 (0) 4 76 87 50 60

*E-mail address:* jean-luc.garden@grenoble.cnrs.fr (J.-L. Garden)

# 1. Introduction

The homopolymer polytetrafluoroethylene (PTFE) has been studied using lots of different physico-chemical methods of investigation [1-3]. Among these experimental methods, thermal analysis or calorimetry is the only one which permits a direct access to thermodynamic parameters such has heat capacity, enthalpy variation, Gibbs free energy variation, entropy variation when a polymer is submitted to a temperature change. For instance, PTFE has been already studied by differential scanning calorimetry (DSC) [4], temperature modulated differential scanning calorimetry (TMDSC) [5], and ac-calorimetry [6, 7]. PTFE has the interesting particularity of undergoing two different physical structural changes at room temperature, one around 292 K and the other around 303 K. It is nowadays admitted that the first transition at 292 K is rather first-order with slow structural changes resulting of the twist of the polymer chains around their symmetry axis [8]. The second transition is seen more as a second-order phase transition with fast kinetic involving molecular scale change due to transition between several conformers and the appearance of disorder along the chain.

However, when these two transitions are studied with dynamical calorimetric methods such as temperature modulated differential scanning calorimetry (TMDSC) or ac-calorimetry, a distinct behaviour is observed between the two transitions [5-7]. The ac-calorimetry method has two advantages: firstly, the frequency can be tuned over a wide range which allows spectroscopic thermal analysis, and secondly this method enables heat capacity measurements with very high sensitivity. Highly resolved heat capacity measurements are currently investigated because it opens up tremendous possible applications in such different fields as the nanophysics [9, 10] or nanobiology [11, 12]. With ac-calorimetry measurements on PTFE, a kinetic effect is directly observed on the heat capacity curve of the first-order transition. Only a few percent of the total enthalpy (*a priori* entirely measured by DSC) is recovered with ac-calorimetry, although on the second transition the two methods give with a good approximation the same results. In the reference [7], a specific calorimetric device has been realized in order to easily vary the oscillating temperature frequency. According to this frequency dependence, a variation of the thermal signature versus frequency has been observed, and a simple physical model has been used to extract a quantitative value of the mean kinetic relaxation time constant of this structural change. The low value of this mean relaxation time as compared to the inverse of the thermal frequency was the explanation of this spectroscopic effect.

In this paper, we study this first-order phase transition under the new point of view that the sample is brought out of thermodynamic equilibrium during this solid-solid transition. We consider for instance that the value of the relaxation time of the process under study is high as compared to the time scale of the experiment bringing then the system in a non-equilibrium thermodynamic state. Consequently, the heat capacity measured over this time scale during the phase transition is the result of a dynamic experiment.

The organization of the paper is as follows:
In the section 2, the calorimetric device used in these ac-calorimetry experiments is thermally fashioned taking into account the non-adiabaticity of the measurement at low frequency and the diffusive regime at high frequency. This allows us to expand the working frequency range of the study. In Sec. 3 and 4, experimental modulus and phase of the modulated temperature of the empty cell and of the loaded cell are provided versus frequency and compared with our thermal model. In Sec. 5, the addenda of the calorimetric device are extracted. Eventually, in Sec. 6 the real part $C'$ and the imaginary part $C''$ of the non-equilibrium complex heat capacity of PTFE are extracted and their thermal variations and frequency dependences are discussed.

## 2. Thermal model of the calorimetric measurement cell

The calorimetric device used for these ac-calorimetry experiments is schematically depicted in the figure 1. The sample is held between two stainless steel thin membranes (12.5 µm thick). On the sides of the metallic membranes not in contact with the sample, thin polyimide films (5 µm thick) are spin coated. A thermometer is micro-patterned on the bottom film, and a heater on the top film, as required for ac calorimetry (see fig. 1). Each stainless steel membrane is glued on a hollow copper cylinder which serves as a constant temperature bath for calorimetric measurements. The top and bottom closures of the copper cylinders are situated few millimetres back of the metallic membranes. The volume between the copper closures and the metallic membrane is connected through a small pipe to a tank of 2 litres filled with gaseous nitrogen under pressure of 1 or 2 bars. This tank which is regulated in temperature is outside the calorimeter. Due to this construction, the thermal link from the metallic membrane to the thermostated bath consists of two parallel terms. The principal one is due to the thermal conductance across the stainless steel membranes and a small one is due to the gas under pressure. The two copper cylinders are tightly clamped on a massive copper piece. This piece is temperature regulated by means of a thermometer (high precision Pt100 resistor) and a heater (high power resistive heater). This piece is thermally linked to a Peltier element which is the cold source of the experiment. The thermometer and the heater are included in a servo-system which allows the temperature of the entire cell to follow temperature ramps, or to be regulated at a constant temperature. The precision is about 0.1 K and the noise is about $10^{-4}\ K/\sqrt{Hz}$. This system is contained in a typical calorimeter enclosure under vacuum with two shields regulated in temperature. For the two micro-patterned sensitive elements (platinum thermometer/ copper-nickel heater) located in the heart of the cell, the leads are brought through the two copper pieces (with specific electrical insulation) till a specific thermal holder regulated in temperature to avoid thermoelectric parasitic effects. The voltage signals are preamplified using home-made low noise preamplifier ($\approx 1\,nV/\sqrt{Hz}$ R.M.S) and then measured by high quality commercial digital voltmeters. The low frequency oscillating current generation chain allows highly resolved thermal power amplitude of typical values between 1 mW and 100 mW ($\Delta P/P \approx 2\Delta V/V \approx 2.10^{-6}$) and low temperature coefficient (~1ppm/K). All these elements gives a heat capacity resolution $\Delta C/C$ of about $\pm 5\times 10^{-6}$. With the absolute value of the total heat capacity of about 20 mJ/K, this calorimeter allows the detection of thermal events as low as ± 100 nJ/K. These efforts made on the electronics of the measurement chain are a necessity for the detection of small imaginary component of the frequency dependent complex heat capacity.

In the basic ac-calorimetry method, a sinusoidal heating power $P_0 \exp(i\omega t)$ is uniformly added to one side of the sample and the temperature response $\delta T_{ac} \exp(i\omega t - \varphi)$ is measured on the other side, the whole being linked to the constant temperature bath by a thermal conductance $K_b$. The measured heat capacity, then, is given by Sullivan and Seidel [14]:

$$|C_{mes}| = \frac{P_0}{\omega \delta T_{ac} \sqrt{1 + \frac{1}{(\omega \tau_1)^2} + (\omega \tau_2)^2 + A}} \tag{1a}$$

and

$$\tan \varphi = f(\omega \tau_1, \omega \tau_2) \tag{1b}$$

$\tau_1 = C/K_b$ with $C$ the heat capacity of the sample. $\tau_2$ is the sum of parasitic contributions of various relaxation times. For instance $\tau_2^2 = \tau_h^2 + \tau_t^2 + \tau_{int}^2$, where $\tau_h$ and $\tau_t$ are the relaxation times of the heater and thermometer towards the sample, respectively. More important is $\tau_{int}$ the internal relaxation time of the sample which is correlated to the diffusivity and the thickness of the sample. $A$ is a constant term depending on $K_b$ and the thermal conductance of the sample. Normally $A << (\omega\tau_2)^2$. $\varphi$ is the phase lag between the input oscillating thermal power and the output resulting temperature modulation.

In this paper, we have considered a thermal model which is more adapted to our calorimetric device. This model is depicted in the figure 2. The calorimetric device is schematically constituted by a multilayered system. More precisely, three different media (1,3,5) with their own heat capacities and thermal conductances are linked to each others by two thermal conductances (media 2, 4) which represent thermal interface between each medium. The medium 1 and the medium 5 are linked to the thermal bath via different thermal conductances $K_{b1}$ and $K_{b2}$. If the values of the internal thermal conductances of each medium and thermal interfaces ($K_{inter2}$ and $K_{inter4}$) are higher than the values of the different heat leaks towards the bath, we can simplified the model assuming that the whole block of different media is linked to the thermal bath by one single thermal conductance $K_b$. In the calorimetric device, the heat leak is thus constituted by the sum of the two thermal conductances ($K_{b1}$ and $K_{b2}$) across the two thin metallic membranes and the small conductance via the gas backward each membrane (see Fig. 1).

As mentioned in the annexe 1, the temperature oscillation $T_3$ measured by the thermometer is calculated using a planar model where the heat diffusion equation is resolved in one-dimensional approach. We also assume that we are in the linear regime where $\delta T_{ac}$ the amplitude of $T_3$ over a period is small as compared to the averaged temperature. As the sample is a composite slab made of five media, we use the matrix method, commonly used in electric circuit theory and clearly explained in the Carslaw and Jaeger book [15]. In this formalism, each medium is represented by a transfer matrix which transforms a vector {heat flow,

temperature} in another vector. Hence, an initial vector {$P_0$, $T_0$} at the position of the heater is transformed via different transfer matrix till the vector {$P_3$, $T_3$} at the level of the thermometer at the end of the medium 5. The equations of the system can be solved if we know the initial conditions for the temperature or the heat flow. The calculus are summarised in the annexe 1. Evidently, in order to have the exact ac-response of the calorimetric device, these calculus have been done assuming that no transition occurs. As in Sullivan and Seidel treatment, the oscillating temperature measured by the thermometer can be separated in a modulus and a phase component, which can be written in a more shortened version:

$$T_3 = \frac{P_0 \exp(i\omega t)}{i\omega(E'-iE'')} \quad (2)$$

where $E'$ and $E''$ are real numbers whose the expressions are given in the annexe 1. This gives rise to a total complex heat capacity where all the components of our calorimetric device are taken into account:

$$C_T = \frac{P_0 \exp(i\omega t)}{i\omega T_3} = E''-iE' \quad (3)$$

In order to obtain the complex heat capacity of the studied sample from this latter equation, we need to know all the thermal conductances and heat capacities of the five media of our model. For that purpose, we have performed a detailed and very precise calibration of our experimental device. In the first measurement there is no PTFE sample in the calorimetric cell, each membrane being clasped together. Then we have measured the cell with a PTFE sample at the centre which is squeezed between the two stainless steel membranes. However, in each case, the thermal interfaces between deposited metallic thin film (heater and thermometer) and the insulation thin polyimide layer have been neglected. In fact, thermal conductance interfaces between deposited metallic thin film and polyimide film of very low heat capacities are so high (few tens of W/K) that they do not play a major role in this thermal model and can be neglected as compared to the other values of the thermal conductances used in this model. Moreover, heat capacities of deposited metallic thin film thinner than 1 µm are very small (less than 1 mJ/K). Hence, relaxation time constants induced by these interfaces are very small as compared to the others taken into account in this study. Moreover, the thermal interfaces between coated thin polyimide insulation layer and the metallic membranes are also neglected for the same reasons. These two different experimental situations are now examined in the next two sections.

**3. Adiabatic plateau and phase behaviour of the empty cell at fixed temperature**

By measuring the empty cell we can determine the heat capacities of the two stainless steel membranes with deposited metallic thin film and polyimide film. We can also get a first value for $K_b$ and the internal conductances.

*3.1 Experiment under the variation of the thermal frequency*

The frequency behaviour of this empty cell is obtained experimentally as a function of the thermal frequency. The experiment has been realized at 283 K, a temperature outside the area of the phase transitions of the PTFE. Two different types of information can be extracted from this experiment.

The first information, called the adiabatic plateau ($\omega\delta T_{ac}$ or $\omega\delta T_{ac}/P_0$ versus $\omega$), is a specific frequency representation of the modulus of the oscillating temperature. This representation (which looks like a plateau) allows the determination of the frequency range where the sample is in perfect thermal equilibrium (not necessarily in thermodynamic equilibrium as we will see in the following when we consider a system where internal degrees of freedom are relaxing). Indeed, at a given frequency the value of the point on the top of the plateau must be directly equals to the inverse of the heat capacity. The experimental plateau measured at 283 K is presented in the figure 3. The second information is inferred when we represent the phase of the modulated temperature versus the frequency. The frequency dependent phase curve is shown in the figure 4. The fits presented in the figures 3 and 4 are issued from our model and are explained in the following sections. These phase and modulus variations are a direct indication that dynamic phenomena due to the non-equilibrium of the sample temperature occur at low and high frequency. At low frequency ($\omega\tau_1<<1$), over one period of the modulation a certain amount of heat relaxes towards the bath via the heat exchange coefficient $K_b$. Somehow, this quantity of heat does not contribute to the heat capacity measurement. At high frequency ($\omega\tau_2>>1$), a part of the input thermal power supplied to the sample does not contribute to the heat capacity measured by the thermometer because of the thermal diffusion within the sample between the heater and the thermometer. The modulus and phase of the measured modulated temperature are simply extracted by Discrete Fourier Transform using Labview software.

*3.2 Fits of the plateau and phase of the empty cell as a function of the thermal frequency.*

In this particular empty cell configuration, the model described in the previous section is used as follows (see fig. 5):

-The medium 1 corresponds to the thin insulating polyimide layer with the thermometer or the heater deposited on the surface.

-The medium 3 corresponds to the two stainless steel membranes.

-The medium 5 is equivalent to the medium 1.

-The two interfaces (medium 2 and 4) form a single medium interface between each metallic membrane.

From this diagram, the calculus made in the annexe 1 give the amplitude and phase of the oscillating temperature at the position of the thermometer. For the fits, numerical values are obtained using the values of the specific heat and thermal conductivity of the stainless steel, the polyimide film and the PTFE sample (used only in the future experiment) which can be found in the literature generally at 298 K [16]. The temperature variations of these parameters have also been taken in the literature in order to have their values at 283 K the temperature of the experiment [17]. The values of these parameters at 298 K and 283 K are shown in table 1. The usual definitions of the heat capacity and the thermal conductance have been used:

$$C = c_p \rho S L \text{ and } K = k \frac{S}{L} \qquad (4)$$

where $c_p$ is the specific heat (J/gK), $\rho$ is the density, $S$ is the surface, $L$ is the length, and $k$ is the thermal conductivity (W/m.K) of the considered materials. In the table 2, the values of the heat capacities, thermal conductances and parameters $\alpha$ (see annexe 1) of the different media 1 and 3 used in the different fits are provided. In order to get accurate fits, three different adjustable parameters have been taken into account:

- the first of these adjustable parameters is the value of the specific heat of the stainless steel at 283 K. A good fit is obtained by taking 0.454 J/gK which is close to the accepted value of 0.44 J/gK found in the literature.

- the second parameter is the value of the thermal link $K_b$ between the device and the thermal bath. If we calculate the value of the horizontal conductance inside the two thin metallic membranes of stainless steel we obtain $K_b \sim 5.67 \times 10^{-3}$ W/K. Furthermore, we have made a dc experiment in order to obtain $K_b$ ($K_b = P_0 / \Delta T_{dc} \sim 4.51 \times 10^{-3}$ W/K). The value of $K_b$ used for the best fit is $5 \times 10^{-3}$ W/K, which is in good agreement with the above values.

- the last of the adjustable parameters are the values $K'_1$ and $K'_2$ the thermal conductance interfaces represented by the medium 2 and 4 (which play the same role in this configuration). In fact, thermal conductance interfaces are very difficult to estimate because they generally depend on different parameters such as the pressure and the surface state. On the figures 3 and 4 two different fits using different values of these thermal interfaces are provided. All the values needed in both fits are shown in the table 2. The first fit is realized with very high values of $K'$, which means in fact that we do not take into account any influence of these interfaces. This fit is not satisfying at high frequencies. The best fit into all the frequency range is obtained with $K'_1 = 0.17$ W/K (for a surface of 1 cm$^2$). The difference between the two fits is due to the low quality of the thermal contact between both stainless steel membranes. (hard/hard contact).

Taking all these parameters into account, the best fit is remarkably adjusted to the experimental plateau and phase curves at different frequencies. Hence, the thermal model used in this configuration may allow us to know very well the thermal behaviour of the calorimetric device, particularly at low and high frequency. At low frequency, spurious frequency effect due to the non adiabaticity of the experiment can be thus corrected. At high frequency spurious frequency effect due to the non homogeneity of the device temperature (diffusion and thermal contact) can be also corrected. Moreover, the heat capacity of the addenda at different frequencies can be extracted from this first type of experiment. Nevertheless, we have to bear in mind that the thermal interface $K'_1$ or $K'_2$ plays a non negligible role in this empty cell experiment. In the following, we will envisage the measuring configuration with a PTFE sample at the centre of the cell and after that explore in details the addenda extraction procedure from these two different configurations.

**4. Adiabatic plateau and phase behaviour of the cell with PTFE sample at fixed temperature.**

In this type of experiment, a 50 μm thick sample of PTFE has been inserted between the two cells. In this configuration, shown in the figure 6, the model is used as follows:

-the medium 1 corresponds to the ensemble polyimide/stainless steel membrane.

-the medium 3 is now constituted by the PTFE sample.

-the media 2 and 4 correspond to the two thermal interfaces $K'_1$ and $K'_2$ between the medium 1 and 3 and 3 and 5 respectively.

-the medium 5 is identical to the medium 1.

In this configuration, the medium 1 is now constituted by a mixture of the previous medium 1 and 3 of the empty cell configuration. Thus, its heat capacity and thermal conductance are simply calculated from the values of the heat capacities and thermal conductances of the previous media 1 and 3. If we write $C_1^0$, $C_3^0$, $K_1^0$ and $K_3^0$ the previous values, then the new values are:

$$C_1 = \frac{1}{2}\left[C_3^0 + 2C_1^0\right] \text{ and } \frac{1}{K_1} = \frac{1}{2}\left[\frac{1}{K_3^0} + \frac{2}{K_1^0}\right] \tag{5}$$

In this experiment, the experimental adiabatic plateau and phase behaviour as a function of the frequency are presented in the figures 7 and 8. The experiment was carried out at a constant temperature of 283 K outside the phase transitions area of the PTFE. On the same figures, two different fits issued from the model are also provided. Once again, the values of the specific heat and thermal conductivity of the different materials and their temperature dependence have been used (see table 1). In the table 3, the values of heat capacities, thermal conductances, $\alpha$ coefficients, and different thermal interfaces necessary for the fits are given. All the parameters used in the empty cell configuration have been kept identical, but new values of $K'_1$ and $K'_2$ must be used. Indeed, once a time the two different fits are calculated, one with high values of the thermal interface and the other with small values. It is interesting to note that in this configuration, the thermal interface does not play such an important role. It is explained by a better thermal interface between PTFE and stainless steel (soft/hard contact) than stainless steel/stainless steel thermal contact (hard/hard contact). In this case, the decrease of the plateau at high frequency is mainly due to the low value of the internal thermal conductance across the thickness of the PTFE sample (see table 3).

Nonetheless, once again the agreement between the model and the experiment is very satisfying. This good agreement between calorimetric device characterization experiments and the thermal model allows sensitive ac-calorimetry experiments over a wide frequency range of about three decades (40 mHz ≤ $F_{th}$ ≤ 40 Hz).

**5. Extraction of the addenda as a function of the temperature**

In the empty cell configuration, an experimental heat capacity curve as a function of the temperature, at a frequency of 1.737 Hz (at the top of the plateau) is provided in the figure 9. Now, using our model in this configuration, taking into account the temperature variations of the different thermodynamic parameters taken from the literature, a temperature dependent heat capacity curve of the addenda is directly extracted and shown in the figure 9.

The temperature dependence of $K_b$ is necessary to obtain this calculated curve; we have simply taken the temperature variation of the thermal conductance of the stainless steel normalized to the value found at 283 K. In other words, the temperature dependence of the small part of the total heat exchange coefficient due to the gas on the back of to the cell has been neglected. These two curves prove that the temperature dependence taken in the literature for the parameters used for the empty cell is correct. In the following, at each experimental frequency, the extracted heat capacity curve as a function of the temperature obtained from the empty cell configuration will be used in order to extract the part of the signal due only to the sample. However, before entering the exact extraction procedure of the frequency dependent complex heat capacity of the PTFE sample during its phase transitions at different frequencies, let recall more precisely the physical model that we used.

## 6. Physical model and extraction of the complex heat capacity

*6.1. Physical model*

In calorimetry, for a given temperature variation, the instantaneous temperature rate defines a time scale of observation. If this time scale is small enough as compared to the kinetic relaxation time constant of a phase transformation, then it is possible that the sample under study is out of its thermodynamic equilibrium state over this time scale. In this case, the measured heat capacity is the result of a non-equilibrium or dynamic experiment [18, 19]. The value of the temperature variation, around a mean constant temperature (stationary condition), can be controlled by the experimentalist via the heater. For example, this results in the choice of the temperature ramp in DSC experiment. In ac-calorimetry, the thermal power is chosen in such a way that the amplitude of the temperature modulation takes a correct value; a correct value is obtained when the amplitude of the temperature modulation is low as compared to the width of the transition. This very intuitive notion is the condition of linearity of the measurement. It is impossible to fulfil this requirement for very weak pure first-order transition where theoretically the transition occurs at a fixed temperature. In the case of configurational changes of PTFE, the phase transitions have a finite width, so the amplitude of the temperature oscillations used in these experiments (approximately 0.26 K) fulfils the condition of linearity. Moreover, in order to verify that the linearity requirement is fulfilled, a good criterion is to measure the "4-ω" term during the experiment. For each experiment, this term was found negligible, which was not the case for amplitude of temperature modulation higher than 1 K. Hence, in the present measurements, we consider that the PTFE is out of thermodynamic equilibrium, but close to equilibrium. Under this condition, when one single relaxation time characterises the transition (Debye relaxation), the dynamic heat capacity can be written:

$$C^* = C' - iC'' = C_\infty + \frac{C_0 - C_\infty}{1 + i\omega\tau} \qquad (6)$$

where

$$C' = C_\infty + \frac{C_0 - C_\infty}{1 + (\omega\tau)^2} \tag{7}$$

is the storage frequency dependent heat capacity, and:

$$C'' = \frac{(C_0 - C_\infty)\omega\tau}{1 + (\omega\tau)^2} \tag{8}$$

is the loss frequency dependent heat capacity [20], where $\omega$ is the angular frequency of the modulated temperature, $C_\infty$ is the heat capacity related to the degrees of freedom of the system having infinitely small relaxation time constants as compared to the inverse of the frequency (generally vibrational modes or phonons bath), and $C_0$ is the total contribution at equilibrium (the frequency is set to zero) of the degrees of freedom having high and small relaxation time constants. The time constant $\tau$ is the kinetic relaxation time constant of the non-equilibrium degree of freedom involved in the transition.

*6.2. Extraction of the real and imaginary part of the frequency dependent complex heat capacity of PTFE.*

Before extracting the real and imaginary parts of the frequency dependent complex heat capacity due only to the thermodynamic behaviour of the PTFE, it is important to take into account the different spurious effects in the complex heat capacity due to the fact that the PTFE is not in thermal equilibrium. The exact thermal model of our device derived in the annexe 1 can be simplified knowing that the frequencies used in these ac-calorimetry measurements are rather low. Indeed, the experimental frequencies never exceed 10 Hz. Thus, the equations (10) and (11) of the annexe 1 giving $E'$ and $E''$ can be simplified to first orders with respect to $\omega$. Hence, from the annexe 1, if we write the complex heat capacity of the PTFE as follows:

$$C_3 = C'_3 - iC''_3 \qquad (9)$$

then a development to the second order in $\omega$ yields to:

$$E'' = C''_3 + \frac{K_b}{\omega} - \omega \left\{ \frac{1}{6}(C'_3 + 2C_1)^2 \left( \frac{1}{K_3} + \frac{2}{K_1} \right) + (C'_3 + C_1) \left[ C_1 \left( \frac{1}{K'_1} + \frac{1}{K'_2} + \frac{1}{3K_3} \right) - \frac{C'_3}{3K_1} \right] \right\} \qquad (10)$$

for the imaginary part of the total complex heat capacity, and:

$$\begin{aligned} E' &= (C'_3 + 2C_1)\left[1 + \frac{K_b}{2}\left(\frac{1}{K_3} + \frac{2}{K_1}\right)\right] + K_b\left[\frac{C'_3}{K'_2} + C_1\left(\frac{1}{K'_1} + \frac{1}{K'_2}\right)\right] \\ &+ \omega C''_3 \left[\frac{C'_3}{3K_3} + C_1\left(\frac{1}{K'_1} + \frac{1}{K'_2} + \frac{1}{K_1} + \frac{1}{K_3}\right)\right] \\ &- \frac{\omega^2}{4}\left\{ \frac{C'^3_3}{30K_3^2} + \frac{4}{3}\frac{C_1^3}{K_1}\left(\frac{1}{K_3} + \frac{4}{5K_1}\right) + \frac{2}{3}C_1^2 C'_3\left(\frac{1}{K_3^2} + \frac{2}{K_1^2} + \frac{4}{K_3 K_1}\right) + \frac{C_1 C'^2_3}{3K_3}\left(\frac{1}{K_3} + \frac{2}{K_1}\right) \right. \\ &\left. + \frac{2}{3}C_1\left(\frac{1}{K'_1} + \frac{1}{K'_2}\right)\left[\frac{C'^2_3}{K_3} + 2\frac{C_1^2}{K_1} + C'_3 C_1\left(\frac{4}{K_1} + \frac{3}{K_3}\right)\right]\right\} \end{aligned} \qquad (11)$$

for the real part of the total complex heat capacity.

The principal aim is now to resolve these two equations in order to obtain $C_3$. For that purpose, it is necessary to know a precise value of $K_b$. Indeed, as we work at low frequency, the heat link has a crucial influence. In order to give a correct estimate for $K_b$, we use experimental data obtained at the smallest possible experimental frequency. Hence, the following treatment is applied only for the two measuring frequencies $F_{th} = 0.06$ Hz and $F_{th} = 0.04$ Hz for which the term $K_b/\omega$ becomes preponderant. We have also additional information which simplifies the resolution of the equations:

- considering only the temperature range where the sample does not undergo any transitions (low and high temperature range) the imaginary part of the complex heat capacity of the sample is obviously equal to zero ($C''_3 = 0$).

Knowing that, in order to have $K_b(T)$, the tangent formed by the two components $E'$ an $E''$ of the complex heat capacity and the modulus of the total complex heat capacity are used. Within the same high and low temperature range and at low frequency, the tangent is written:

$$\tan \varphi = \frac{E'}{E''} \approx \frac{(C'_3+2C_1)\left[1+\frac{K_b}{2}\left(\frac{1}{K_3}+\frac{2}{K_1}\right)\right]+\left(\frac{K_b C_1}{K'_1}\right)}{\frac{K_b}{\omega}-\omega\left\{\frac{1}{6}(C'_3+2C_1)^2\left(\frac{1}{K_3}+\frac{2}{K_1}\right)+(C'_3+C_1)\left[C_1\left(\frac{1}{K'_1}+\frac{1}{3K_3}\right)-\frac{C'_3}{3K_1}\right]\right\}} \quad (12)$$

For the sake of simplicity we write this expression as:

$$\tan \varphi = \frac{E'}{E''} \approx \frac{(C'_3+2C_1)A}{\frac{K_b}{\omega}-B\omega} \quad (13)$$

and for the modulus of the total complex heat capacity:

$$\left|\frac{P_0}{\omega T_3}\right| = \left(\frac{K_b}{\omega}-B\omega\right)\sqrt{1+\tan^2 \varphi} \quad (14)$$

with

$$A = 1 + K_b \left[ \frac{1}{2} \left( \frac{1}{K_3} + \frac{2}{K_1} \right) + \frac{C_1}{K'_1 (C'_3 + 2C_1)} \right] \tag{15}$$

and

$$B = \left\{ \frac{1}{6} (C'_3 + 2C_1)^2 \left( \frac{1}{K_3} + \frac{2}{K_1} \right) + (C'_3 + C_1) \left[ C_1 \left( \frac{1}{K'_1} + \frac{1}{K'_2} + \frac{1}{3K_3} \right) - \frac{C'_3}{3K_1} \right] \right\} \tag{16}$$

As $A$ and $B$ are second order terms, we can calculate them by substituting for $K_b$ and $C'_3$ in the equations (15) and (16) their first order approximations. From the last expressions of the modulus and phase at low frequency, it is now possible to extract the temperature variation of $K_b$ from the experimental measured curves at the frequencies of 0.06 Hz and 0.04 Hz in the range of temperature outside of the transitions:

$$K_b = \omega \left[ \frac{\left| \frac{P_0}{\omega T_3} \right|}{\sqrt{1 + \tan^2 \varphi}} + B\omega \right] \tag{17}$$

On the figure 10, the temperature dependence of $K_b$ resulting from this previous treatment is provided. We used a single extrapolation between the data obtained at T < 288 K and T > 308 K for covering all the temperature range. On the same figure, the temperature dependence of the stainless steel membranes thermal conductance is also shown for comparison. There is a pretty good agreement between both curves; the difference could be due to some various thermal dependence of the thermal conductivity of the stainless steel as observed in the literature. Even if they are not represented on the figure, it has to be noted that the extracted values of $K_b(T)$ are rather identical for the two different frequencies, which validates our approach of extracting the temperature behaviour of $K_b$.

Now $K_b$ is well-known, so we can resolve by successive approximations the equations (10) and (11) and extract precise values for $C'_3$ and $C''_3$. From the equation (10) and (11) the first order term of $C''_3$ and $C'_3$ are:

$$C''_3 = E'' - \frac{K_b}{\omega} \qquad (18)$$

and

$$C'_3 = E' - 2C_1 \qquad (19)$$

The equation (11) can be rewritten as an equation of second degree in $C'_3$:

$$\begin{aligned}
0 = {} & C'^2_3 \frac{\omega^2}{2} \frac{C_1}{K_3} \left( \frac{1}{6K_3} + \frac{1}{3K_1} + \frac{1}{3K'_1} + \frac{1}{3K'_2} \right) \\
& + C'_3 \left\{ -1 - \frac{K_b}{K'_2} - \frac{\omega C''_3}{3K_3} + \frac{\omega^2 C_1^2}{6} \left[ \frac{1}{K_3^2} + \frac{2}{K_1^2} + \frac{4}{K_3 K_1} + \left( \frac{1}{K'_1} + \frac{1}{K'_2} \right) \left( \frac{4}{K_1} + \frac{3}{K_3} \right) \right] \right\} \\
& + E' - C_1 \left[ 2 + K_b \left( \frac{1}{K'_1} + \frac{1}{K'_2} \right) + \omega C''_3 \left( \frac{1}{K'_1} + \frac{1}{K'_2} + \frac{1}{K_1} + \frac{1}{K_3} \right) - \frac{\omega^2 C_1^2}{3K_1} \left( \frac{1}{K'_1} + \frac{1}{K'_2} + \frac{4}{5K_1} + \frac{1}{K_3} \right) \right]
\end{aligned} \qquad (20)$$

where the small third order term, $\frac{\omega^2}{120}\frac{C_3'^3}{K_3^2}$ has been neglected[1]. The resolution of this second order equation gives $C'_3$ at the second order. Eventually, with $C'_3$ calculated from equation (20), we obtain $C''_3$ of second order.

Before giving the results of the previous extracting procedure, we present on the figures 11 and 12 the modulus and phase of the measured total frequency dependent complex heat capacity (raw data) as a function of the temperature at three well-distinct frequencies. One is presented at low frequency (0.04 Hz), the second is located on the adiabatic plateau (0.4 Hz), and the last at rather high frequency (4 Hz). The experimental conditions used for these ac-calorimetry experiments are:

- the sample is a 50 μm thick PTFE film purchased at GoodFellow.

- the temperature amplitude has been maintained at a constant value for all the frequencies and it is of the order of 0.26 K peak to peak. The amplitude of the thermal power supplied has been adjusted to fulfil this requirement. This constant temperature amplitude allows the linearity condition to be fulfilled for all the experiments realized at different frequencies.

- in the case of measurements versus the temperature at fixed frequency, the dc temperature has followed a slow linear ramp with a rate of 0.3 K/min excepted for the two low frequencies of 0.04 Hz and 0.06 Hz for which the ramp was 0.1 K/min. These small values of the temperature rate allow the stationary criteria to be fulfilled. Indeed, with the usual ramp used (0.3 K/min) we have observed that at these low frequencies, the ramp has a small influence on the magnitude of the peak. The decrease of the ramp by a factor 3 (to 0.1 K/min) gives now $dT/dt \approx 1.7 mK/s$ and for the lowest frequency, $4F_{th}\delta T_{ac} \approx 21 mK/s$. More than a factor of 10 between the two temperature rates seems to be a good criterion in order to respect the stationary condition. In this kind of experiment, the highest value of the frequency is only 10 Hz.

---

[1] To be more precise, this small term (second order in ω) has been added to the constant term of the equation (20), avoiding the resolution of an equation of the third degree. In this case, $C'_3$ has been taken from equ. (19).

- the temperature range has been spread between approximately 278 K and 323 K.

- the sample has been maintained during one night at about 278 K before each temperature experiment.

For the low and high frequencies, the modulus and phase are deformed by the spurious complex components due to the experimental conditions of the measurement and the influence of the device, which are, as already mentioned, the adiabaticity and temperature homogeneity requirements.

Once the corrections have been made following the previous extracting procedure, it is then possible to directly extract from these curves the two components $C'_3$ and $C''_3$ of the frequency dependent complex heat capacity of the PTFE. On the figures 13 and 14, the temperature dependence of $C'_3$ and $C''_3$ is presented at these three different frequencies. The discussion of both figures will be separated in two parts. In the first part, we will analyze the data obtained in the temperatures outside the temperature range of the two phase transitions. The second part will be devoted to the discussion of the data obtained in the temperature range where PTFE undergoes two phase transitions.

- T < 288 K and T > 308 K: in this temperature range, we are expecting that due to very small frequency effects the $C'_3(T)$ are identical whatever the frequency. We have obviously the same behaviour for the $C''_3(T)$ curves. Concerning the real part $C'_3$, it is exactly what we observe excepted for the curve at 0.04 Hz in the high temperature range. By now, we think that this effect could be due to some phase calibration problems. For the imaginary part $C''_3(T)$, the agreement is good for T less than 288 K, but in the high temperature range, higher is the frequency bigger is the discrepancy. It is likely that the problem is due to the temperature dependence of the thermal conductivity of the PTFE. The thermal conductance $K_3(T)$ we have used in our calculations is a continuous curve as a function of the temperature. Due to the phase transitions, $K_3(T)$ could undergo some step-like variation at 292 K and 303 K. Moreover, the second order value of $C''_3(T)$ is depending of an $\omega$ term which is itself function of $K_3(T)$.

- 288 K < T < 308 K: firstly, we notice that the value of the imaginary part $C''_3(T)$ is much more smaller than the value of the real part. This is not surprising if we have in mind that this imaginary part is directly associated to the mean entropy production due to the relaxation of some internal degrees of freedom of the PTFE averaged over the time scale of the experiment (one cycle of the temperature oscillation). Indeed, this entropy creation keeps small in the vicinity of equilibrium [19]. Secondly, a frequency dependence is seen on the two components of the complex heat capacity. For the real part, the decrease of the peak when the frequency is increased represents the freezing of internal degrees of freedom involved in the phase transition when it is observed under different time scales [19, 21]. For the imaginary part, as we have already mentioned, it corresponds to the decrease of the net entropy produced during the irreversible process occurring within the PTFE when it undergoes its phase transitions. We remark that over this frequency range, this entropy production decreases when the frequency increases. This indicates that the typical relaxation time constant involved in the phase transition is certainly higher than the different observation time scales used in these experiments. Thirdly, a second peak is visible on $C''_3$, which seems to indicate that irreversible effects occur also for the second phase transition of the PTFE around 303 K. This aspect is still under study.

On the figure 15, the Cole-Cole plot $C''_3 = f(C'_3)$ is presented at a temperature around the maximum of the first peak where the sensitivity to the frequency variation is maximum. On the same figure we have shown a Cole-Cole plot directly issued from the simple Debye model with a value of $\tau$ equal to 10 s (equation 6). We have to bear in mind that the use of different values of $\tau$ does not modify the curvature of a Cole-Cole plot. It is clearly seen that this model does not fit the experimental results. Consequently, the fact that one single internal degree of freedom with a specific kinetic relaxation time constant is involved in this phase transition is not judicious. A physical model including a distribution of relaxation time constants could be more appropriate, but this goes beyond the scope of this article.

As an interesting indication, on the last figure 16 and 17, the real part $C'_3$ and the imaginary part $C''_3$ of the PTFE are shown as a function of the frequency at few different temperatures. Two remarks can be made:

-no kinetic effects are visible outside the transition zone.

-the typical kinetic relaxation time constant of the PTFE undergoing the 292 K phase transition has a value higher than the biggest value of the period of the thermal oscillation used in these ac-calorimetry experiments. At this level, certainly the temperature modulated differential scanning calorimetry experiments (TMDSC) could be of great interest because the period of the temperature oscillation used in this type of experiment are generally much longer than in ac-calorimetry experiments [5].

The crystallized phase of the PTFE sample used in our experiments ranges between 50% and 99% as indicated by GoodFellow company. The combined heat of the two phase transformations that we have obtained by DSC on this sample (temperature rate of 0.2 K/min) is around 6.8 J/gK [7], a value which is in good agreement with [5]. In accordance with [24], this value indicates a degree of crystallinity ranging between 52% and 67%. However, as Androsch seems to suggest in the reference [5], qualitatively the thermal history and crystallization conditions of PTFE should not affect significantly the transition kinetics. Subsequently, a qualitative comparison on PTFE with these two different calorimetric methods is possible. Futatsugi and *al.* have first recognized that a kinetic effect is clearly visible in the first phase transition of the PTFE, when the heat capacity is measured by ac-calorimetry [6]. With the TMDSC method, Androsch first demonstrates that it is possible to observe more precisely this kinetic effect when varying the thermal frequency. In the figure 2 of the reference [5], a decrease of the amplitude of the heat capacity peak is clearly observed when the frequency is increased in a good agreement with our results. Moreover, at the lowest amplitude of modulation (0.1 K), in the reference [5] the smallest peak which corresponds to a frequency of 16.7 mHz is higher than the highest peak of our ac-calorimetry measurement obtain for a frequency of 40 mHz. The effect of the amplitude of the modulation on the amplitude of the heat capacity peak observed in the figure 2 of reference [5] is certainly due, as Androsch himself indicates, to the stationary criterion which is not respected because the dc temperature ramp rate ($dT/dt$) is of the same order as the ac temperature rate ($\approx 4\delta T_{ac} F_{th}$). In our case, as explained previously we are in stationary condition. In fact, the most important progress that has been made in the present paper is to provide for the first time on the PTFE, the real and imaginary components of the frequency dependent heat capacity during the low temperature phase transitions of this polymer.

## 7. Conclusion

In this paper, we have presented a thermodynamic investigation of a polymer, the PTFE, which undergoes two phase transitions at room temperature. The heat capacity of the PTFE has been measured between 278 K and 323 K by means of the ac-calorimetry method.

The device used for the experiments has been built up by means of micro-fabrication technology. Usually, since the work of Sullivan and Seidel, the device and the sample are assumed to constitute one single diffusive media with thermal contact between sensitive elements and the sample, the whole being thermally anchored to a thermal bath via a determined heat leakage. In our case, the calorimetric device holding the sample is better thermally represented by a multi-layered slab of different heat capacities, thermal conductances and interfaces. Thus, the temperature modulation measured by the thermometer has been derived using this model.

The oscillating temperature is described by a modulus and a phase which are together temperature and frequency dependent. Following our model, this gives rise to a measured heat capacity which can be represented by a complex number with a temperature and frequency dependent real and imaginary component.

From the experimental data in the adiabatic plateau the modulus and phase of this modulated temperature have been presented as a function of the frequency and compared to different fits following the model. Thanks to an adequate agreement between theory and experiments, it has been then possible to enlarge the interesting frequency range where *a priori* the heat capacity of the sample is frequency independent. As our system is not differential, we have proceeded to empty cell experiments in order to know exactly the temperature and frequency behaviour of the heat capacity of the addenda. These experiments have been next followed by experiments with loaded cell as a function of the temperature over a frequency range located between 0.04 Hz and 10 Hz.

From these experiments, the real and imaginary part of the measured complex heat capacity of the PTFE has been extracted from our model in temperature and in frequency. To our knowledge, it is the first time that the real and imaginary parts of the complex heat capacity are presented in such a ac-calorimetry experiments on solid-solid phase transition in polymer [22, 23]. The two components exhibit a frequency dependence clearly seen on the first phase transition of the PTFE at 292 K. On a macroscopic non-equilibrium thermodynamic point of view, this effect indicates that depending on the time scale of observation (one period of the temperature modulation) the sample is outside equilibrium with respect to one of its internal degrees of freedom involved in the phase transformation. In the high temperature range, a difference exists between curves extracted at various frequencies which should not be the case. It is certainly due to the variation of the thermal conductivity of the PTFE during the two phase transitions, this variation being not taken into account in our model. Then, the two complex heat capacity components have been provided as a function of the frequency at different temperatures. The decrease of the imaginary part when the frequency increases clearly indicates that the characteristic kinetic time constant involved in the first phase transition of the PTFE should be much larger than the highest period of temperature oscillations used in the experiments. A Cole-Cole plot of $C'' = f(C')$ has also been presented and compared to the Debye model with one single time constant. Unfortunately, the discrepancy between experiments and Debye relaxation is important.

In a future work, we will envisage some new directions:

- a change of the thermal conductivity of the PTFE due to the phase transitions will be incorporate in our model in order to obtain identical temperature dependence for different frequencies.

- the unexpected $C''$ existence for the second phase transition of the PTFE at 303 K will be investigate in more details.

- the Debye model will be abandoned, a distribution of kinetic time constants will be considered yielding to a slightly modified expression of the frequency dependent complex heat capacity.

- although they are different in their basic foundations, it may be interesting to compare low frequency ac-calorimetry experiments with high frequency TMDSC experiments for a same sample at same frequencies in order to improve comprehension of non-equilibrium phase transition in PTFE or other polymers.

This work was realized inside the team of Thermodynamique des Petits Systèmes and the Pôle de Capteurs Thermométriques et Calorimétrie of the Institut Néel. The authors want to thank E. Château, P. Lachkar, E. André; J.-L. Bret, C. Guttin, P. Brosse-Maron. We are infinitely indebted to J. Chaussy without whom this work would have not been possible.

**Annexe1: <u>Calculus of the modulated temperature of a sample constituted by five different media.</u>**

Let us come back to the thermal drawing of the figure 2. The sample under calorimetric investigations is constituted by five different media whose two of them are only thermal interfaces. The whole is thermally linked to a thermal bath via a small thermal conductance $K_b$, which is represented by the sum of the different heat links (because this conductance is smaller than the internal thermal interfaces). The symmetry of the problem allows us to work with one single dimension represented by the coordinate $x$ which characterizes the direction of the flow of heat. The thermal power $P$, which is the sum of a dc and an ac term, is directly supplied to the top face of the medium 1. Indeed we assume that the heater is placed on this face at the x = 0 coordinate. The temperature/heat flux couple is written at this coordinate $(T_0, P_0)_{x=0}$. Our principal aim is to calculate the temperature/heat flux couple $(T_3, P_3)_{x=L}$ at the bottom face of the last medium 5, where the thermometer is present. Indeed, it is only at this distance $L$ from the heater that the modulus and phase of the oscillating temperature is measured by the thermometer.

The method of propagation of heat used in this paper is the matrix method used by Carslaw and Jeager [15]. The thermal path within the medium i is represented by a matrix written $A_i$. For the sake of symmetry, the medium 5 is equivalent to the medium 1 ($A_1 = A_5$). Hence, the temperature/heat flux couple $(T_3, P_3)_{x=L}$ is relied to the temperature/heat flux couple $(T_0, P_0)_{x=0}$ as the following:

$$\begin{bmatrix} T_3 \\ Q_3 \end{bmatrix} = A \begin{bmatrix} T_0 \\ Q_0 \end{bmatrix} = \begin{bmatrix} a_1 & b_1 \\ c_1 & d_1 \end{bmatrix} \begin{bmatrix} 1 & -(1/K'_2) \\ 0 & 1 \end{bmatrix} \begin{bmatrix} a_3 & b_3 \\ c_3 & d_3 \end{bmatrix} \begin{bmatrix} 1 & -(1/K'_1) \\ 0 & 1 \end{bmatrix} \begin{bmatrix} a_1 & b_1 \\ c_1 & d_1 \end{bmatrix} \begin{bmatrix} T_0 \\ Q_0 \end{bmatrix} \qquad (1)$$

where $A = A_1 A_4 A_3 A_2 A_1$ is the product of the transfer matrix of the different media.

For the heat flux, the boundary conditions are simply: $Q_0 = P_0$ and $Q_3 = K_b T_3$ .(2)

If we write $A = \begin{bmatrix} a & b \\ c & d \end{bmatrix}$, where $a$, $b$, $c$, $d$ are complex numbers, then we have:

$$T_3 = \frac{P_0(ad - bc)\exp(i\omega t)}{aK_b - c} \tag{3}$$

where the $a_j$, $b_j$, $c_j$, $d_j$ are given by:

$$\begin{cases} a_j = ch\theta_j & b_j = -\frac{sh\theta_j}{K_j\theta_j} \\ c_j = -K_j\theta_j sh\theta_j & d_j = c_j \end{cases} \quad \text{for } (j = 1, 3) \tag{4}$$

with $\theta_j = \sqrt{\frac{\omega}{2D_j}}(1+i)L_j = (1+i)\alpha_j$ and $\alpha_j = \sqrt{\frac{\omega}{2D_j}}L_j = \sqrt{\frac{\omega C_{0j}}{2K_j}}$ (5)

with $K_j = k_j \dfrac{S_j}{L_j}$ and $D_j = \dfrac{k_j}{\rho_j c_{0j}}$ and $C_{0j} = \rho_j S_j L_j c_{0j}$ (6)

with $k_j$ the thermal conductivity, $\rho_j$ the density and $c_{0j}$ the specific heat of the medium $j$.

For the calculus of the component of the total transfer matrix we use the fact that:

$$\det A = ad - bc = \det(A_1)\det(A_4)\det(A_3)\det(A_2)\det(A_1) = 1 \tag{7}$$

Then:

$$T_3 = \frac{P_0 \exp(i\omega t)}{aK_b - c} \tag{8}$$

where *a* and *c* are calculated using the equations (1), (4), (5) and (6).

The resulting temperature at a distance *L* from the heater can be written:

$$T_3 = \frac{P_0 \exp(i\omega t)}{i\omega(E' - iE'')} \tag{9}$$

where *E'* and *E''* are real numbers given by the two cumbersome expressions:

$$\begin{aligned}
E' =& \frac{C_{01}}{2\alpha_1}\left[sh2\alpha_1 \cos 2\alpha_1(ch\alpha_3 \cos\alpha_3 + sh\alpha_3 \sin\alpha_3) - ch2\alpha_1 \sin 2\alpha_1(sh\alpha_3 \sin\alpha_3 - ch\alpha_3 \cos\alpha_3)\right] \\
&+ \frac{C_{03}}{4\alpha_3}\left[(sh\alpha_3 \cos\alpha_3 + ch\alpha_3 \sin\alpha_3)(1 + \cos 2\alpha_1 ch2\alpha_1) + sh2\alpha_1 \sin 2\alpha_1(sh\alpha_3 \cos\alpha_3 - ch\alpha_3 \sin\alpha_3)\right] \\
&+ \frac{C_{01}^2 \alpha_3}{4C_{03}\alpha_1^2}\left[(sh\alpha_3 \cos\alpha_3 + ch\alpha_3 \sin\alpha_3)(-1 + \cos 2\alpha_1 ch2\alpha_1) + sh2\alpha_1 \sin 2\alpha_1(sh\alpha_3 \cos\alpha_3 - ch\alpha_3 \sin\alpha_3)\right] \\
&+ \frac{K_b}{\omega}\left(sh\alpha_3 \sin\alpha_3 ch2\alpha_1 \cos 2\alpha_1 + ch\alpha_3 \cos\alpha_3 sh2\alpha_1 \sin 2\alpha_1\right) \\
&+ \frac{K_b}{2\omega}\left(\frac{C_{03}\alpha_1}{C_{01}\alpha_3} + \frac{C_{01}\alpha_3}{C_{03}\alpha_1}\right)\left[ch\alpha_3 \sin\alpha_3 sh2\alpha_1 \cos 2\alpha_1 + sh\alpha_3 \cos\alpha_3 ch2\alpha_1 \sin 2\alpha_1\right] \\
&+ \left(\frac{1}{K'_1} + \frac{1}{K'_2}\right)\left\{\begin{array}{l}\frac{\sqrt{C_{03}K_3 C_{01}K_1}}{2}\left[sh\alpha_3 \cos\alpha_3 sh2\alpha_1 \cos 2\alpha_1 - ch\alpha_3 \sin\alpha_3 ch2\alpha_1 \sin 2\alpha_1\right] \\ + \frac{C_{01}K_1}{2}\left[(ch2\alpha_1 \cos 2\alpha_1 ch\alpha_3 \cos\alpha_3 - sh\alpha_3 \sin\alpha_3 sh2\alpha_1 \sin 2\alpha_1) - ch\alpha_3 \cos\alpha_3\right]\end{array}\right\} \\
&+ \left(\frac{\alpha_3 K_3 K_1 C_{01}}{2K'_1 K'_2}\right)\left[(ch2\alpha_1 \cos 2\alpha_1 - 1)(sh\alpha_3 \cos\alpha_3 - ch\alpha_3 \sin\alpha_3) - sh2\alpha_1 \sin 2\alpha_1(sh\alpha_3 \cos\alpha_3 + ch\alpha_3 \sin\alpha_3)\right] \\
&+ \frac{K_b}{\omega}\left(\frac{1}{K'_1} + \frac{1}{K'_2}\right)\left\{\begin{array}{l}\frac{K_1\alpha_1}{2}\left[\begin{array}{l}(ch\alpha_3 \cos\alpha_3 - sh\alpha_3 \sin\alpha_3)ch2\alpha_1 \sin 2\alpha_1 \\ + (sh\alpha_3 \sin\alpha_3 + ch\alpha_3 \cos\alpha_3)sh2\alpha_1 \cos 2\alpha_1\end{array}\right] \\ + \frac{K_3\alpha_3}{2}\left[\begin{array}{l}(ch\alpha_3 \sin\alpha_3 + sh\alpha_3 \cos\alpha_3)(\cos 2\alpha_1 ch2\alpha_1 - 1) \\ + (sh\alpha_3 \cos\alpha_3 - ch\alpha_3 \sin\alpha_3)sh2\alpha_1 \sin 2\alpha_1\end{array}\right]\end{array}\right\} \\
&+ \frac{K_b}{\omega}\frac{K_3 K_1}{K'_1 K'_2}\alpha_3\alpha_1\left[sh\alpha_3 sh2\alpha_1 \cos\alpha_3 \cos 2\alpha_1 - ch\alpha_3 ch2\alpha_1 \sin\alpha_3 \sin 2\alpha_1\right] \\
&+ \frac{K_b}{\omega}\frac{K_3\alpha_3}{K'_2}(ch\alpha_3 \sin\alpha_3 + sh\alpha_3 \cos\alpha_3)
\end{aligned} \tag{10}$$

and

$$\begin{aligned}
E'' = &\frac{C_{01}}{2\alpha_1}\left[sh2\alpha_1\cos 2\alpha_1(ch\alpha_3\cos\alpha_3 - sh\alpha_3\sin\alpha_3) - ch2\alpha_1\sin 2\alpha_1(sh\alpha_3\sin\alpha_3 + ch\alpha_3\cos\alpha_3)\right] \\
&+ \frac{C_{03}}{4\alpha_3}\left[(sh\alpha_3\cos\alpha_3 - ch\alpha_3\sin\alpha_3)(1+\cos 2\alpha_1 ch2\alpha_1) - sh2\alpha_1\sin 2\alpha_1(sh\alpha_3\cos\alpha_3 + ch\alpha_3\sin\alpha_3)\right] \\
&+ \frac{C_{01}^2\alpha_3}{4C_{03}\alpha_1^2}\left[(sh\alpha_3\cos\alpha_3 - ch\alpha_3\sin\alpha_3)(-1+\cos 2\alpha_1 ch2\alpha_1) - sh2\alpha_1\sin 2\alpha_1(sh\alpha_3\cos\alpha_3 + ch\alpha_3\sin\alpha_3)\right] \\
&+ \frac{K_b}{\omega}\left(ch\alpha_3\cos\alpha_3 ch2\alpha_1\cos 2\alpha_1 - sh\alpha_3\sin\alpha_3 sh2\alpha_1\sin 2\alpha_1\right) \\
&+ \frac{K_b}{2\omega}\left(\frac{C_{03}\alpha_1}{C_{01}\alpha_3} + \frac{C_{01}\alpha_3}{C_{03}\alpha_1}\right)\left[sh\alpha_3\cos\alpha_3 sh2\alpha_1\cos 2\alpha_1 - ch\alpha_3\sin\alpha_3 ch2\alpha_1\sin 2\alpha_1\right] \\
&- \left(\frac{1}{K'_1}+\frac{1}{K'_2}\right)\left\{\begin{array}{l}\frac{\sqrt{C_{03}K_3 C_{01}K_1}}{2}\left[ch\alpha_3\sin\alpha_3 sh2\alpha_1\cos 2\alpha_1 + sh\alpha_3\cos\alpha_3 ch2\alpha_1\sin 2\alpha_1\right] \\ +\frac{C_{01}K_1}{2}\left[(sh2\alpha_1\sin 2\alpha_1 ch\alpha_3\cos\alpha_3 + sh\alpha_3\sin\alpha_3 ch2\alpha_1\cos 2\alpha_1) - sh\alpha_3\sin\alpha_3\right]\end{array}\right\} \\
&- \left(\frac{\alpha_3 K_3 K_1 C_{01}}{2K'_1 K'_2}\right)\left[(ch2\alpha_1\cos 2\alpha_1 - 1)(sh\alpha_3\cos\alpha_3 + ch\alpha_3\sin\alpha_3) + sh2\alpha_1\sin 2\alpha_1(sh\alpha_3\cos\alpha_3 - ch\alpha_3\sin\alpha_3)\right] \\
&+ \frac{K_b}{\omega}\left(\frac{1}{K'_1}+\frac{1}{K'_2}\right)\left\{\begin{array}{l}\frac{K_1\alpha_1}{2}\left[\begin{array}{l}(ch\alpha_3\cos\alpha_3 - sh\alpha_3\sin\alpha_3)sh2\alpha_1\cos 2\alpha_1 \\ -(sh\alpha_3\sin\alpha_3 + ch\alpha_3\cos\alpha_3)ch2\alpha_1\sin 2\alpha_1\end{array}\right] \\ +\frac{K_3\alpha_3}{2}\left[\begin{array}{l}(sh\alpha_3\cos\alpha_3 - ch\alpha_3\sin\alpha_3)(\cos 2\alpha_1 ch2\alpha_1 - 1) \\ -(ch\alpha_3\sin\alpha_3 + sh\alpha_3\cos\alpha_3)sh2\alpha_1\sin 2\alpha_1\end{array}\right]\end{array}\right\} \\
&- \frac{K_b}{\omega}\frac{K_3 K_1}{K'_1 K'_2}\alpha_3\alpha_1\left[ch\alpha_3 sh2\alpha_1\sin\alpha_3\cos 2\alpha_1 + sh\alpha_3 ch2\alpha_1\cos\alpha_3\sin 2\alpha_1\right] \\
&+ \frac{K_b}{\omega}\frac{K_3\alpha_3}{K'_2}(sh\alpha_3\cos\alpha_3 - ch\alpha_3\sin\alpha_3)
\end{aligned}$$

(11)

Fig. 1 The schematic view of the ac calorimeter. The central zone is zoomed with more details. The PTFE sample is directly in contact with the two stainless-steel membranes. On the other side of the membranes, the heater and the thermometer have been micro-patterned on a polyimide insulation layer which has been spin-coated on each membrane.

Fig. 2 A general thermal model system used for the exact calculation of the heat capacity of the PTFE sample $C_3$. The heater and the thermometer are evaporated onto the medium 1 and 5 respectively. We neglect their heat capacity and their thermal contact resistances. Thermal conductances are represented. In this drawing, the dimensions of the different parts are not respected.

Fig.3 The modulus of the inverse apparent heat capacity of the empty cell as a function of the frequency at 283 K. The points are experimental data. The full and broken lines are obtained from fits with equation (9) in the annexe 1. The fitting parameters are shown in table 2.

Fig.4 Frequency dependence of the phase shift between the heater and the thermometer plus 90°, for an empty cell. We added 90° in order to have a zero phase shift on the « plateau ». The points are experimental data. The full and broken lines are obtained from fits with equation (9) in the annexe 1. The fitting parameters are shown in table 2.

Fig.5 The model of the empty cell. In this model we made the same hypothesis for the heater and thermometer than in the Fig.2. We also suppose perfect contacts between the media 1 and 3 and between the media 5 and 1. Nevertheless, the thermal interface between the media 3 and 5 is taken into account by the medium 2. Here $K_b$ is the total thermal conductance for all the heat links between the different media and the thermal bath.

Fig 6 The model of the cell filled with a PTFE sample. Here the medium 1 is made of the media 1 and 3 of the empty cell.

Fig 7 The modulus of the inverse apparent heat capacity of the cell, filled with a PTFE sample, as a function of the frequency at 283 K. The points are experimental data. The full and broken lines are obtained from fits with equation (9) in the annexe 1. The fitting parameters are shown in table 3

Fig.8 Frequency dependence of the phase shift between the heater and the thermometer plus 90°, for a cell filled with a PTFE sample. We added 90° in order to have a zero phase shift on the « plateau ». The points are experimental data. The full and broken lines are obtained from fits with equation (9) in the annexe 1. The fitting parameters are shown in table 3

Fig. 9 Temperature dependence of the heat capacity of the total empty cell made at the fixed frequency of 1,737 Hz. The points represent the experimental data and the full line the results of a fit using equation (9) in annexe 1 with the fitting parameters shown in the table 4.

Fig 10. Temperature dependence of $K_b$, the thermal conductance of the total heat link between the different media and the thermal bath. The solid line for $K_b^1$ is the result of the measurements at 0.04Hz. Using equation (17) we can extract $K_b$. The broken line for $K_b^2$ is derived, after a normalisation made at 283 K (see text), with the use of the thermal variation of the stainless steel thermal conductivity.

Fig.11 Temperature dependence of the modulus of the apparent heat capacity for three different frequencies: 0.04 Hz, 0.4 Hz and 4Hz. We can observe the frequency effect on the first peak around 294 K.

Fig 12 Temperature dependence of the phase (plus 90°) of the apparent heat capacity for three different frequencies : 0.04 Hz, 0.4 Hz and 4Hz.

Fig.13 The real part of the specific heat of a PTFE sample ($C'_3$) as a function of the temperature for three different frequencies: 0.04 Hz, 0.4 Hz and 4Hz.

Fig.14 The imaginary part of the specific heat of a PTFE sample ($C''_3$) as a function of the temperature for three different frequencies: 0.04 Hz, 0.4 Hz and 4Hz.

Fig.15 Cole–Cole plot of the imaginary part versus the real part of the specific heat of a PTFE sample made at a fixed temperature T=294 K. The points are experimental data, the solid line is a fit of a Debye model with $\tau$ =10s.

Fig. 16 The real part of the specific heat of a PTFE sample as a function of the frequency for five different temperatures: T=283,290,294,296,308 K.

Fig. 17 The imaginary part of the specific heat of a PTFE sample as a function of the frequency for five different temperatures: T=283,290,294,296,308 K.

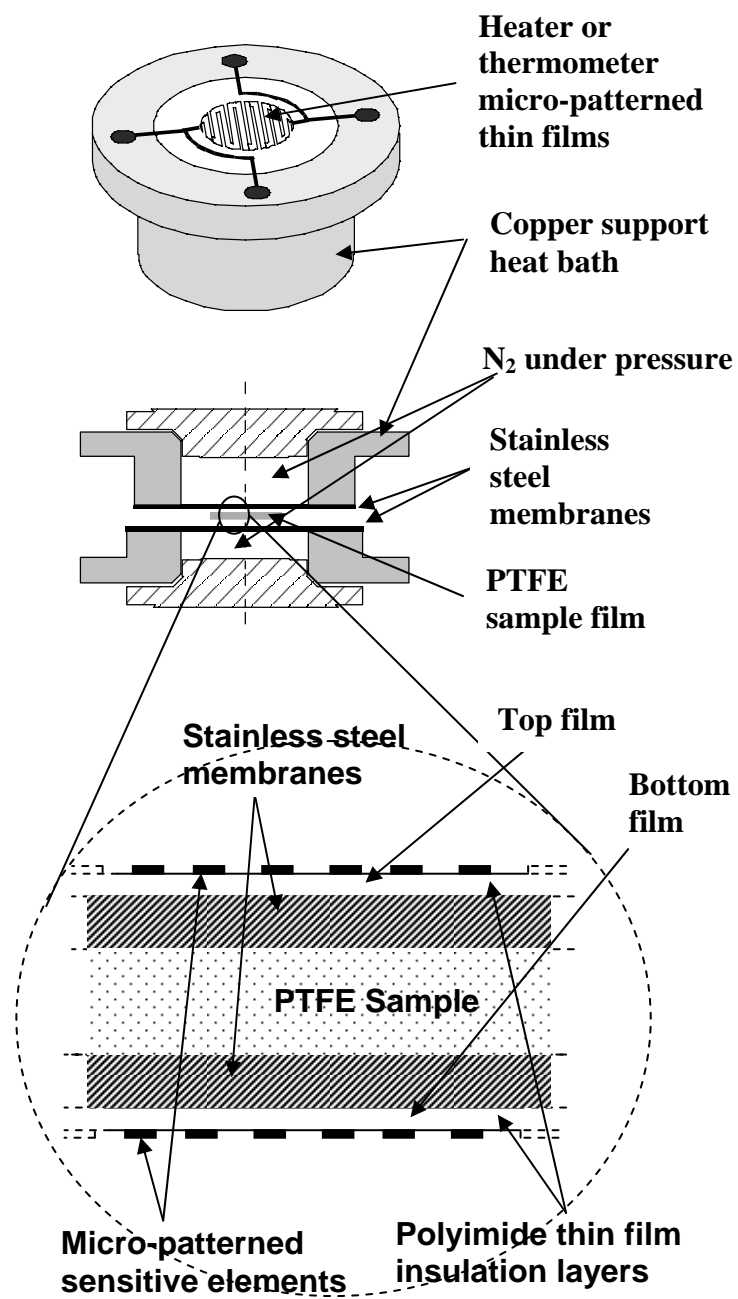

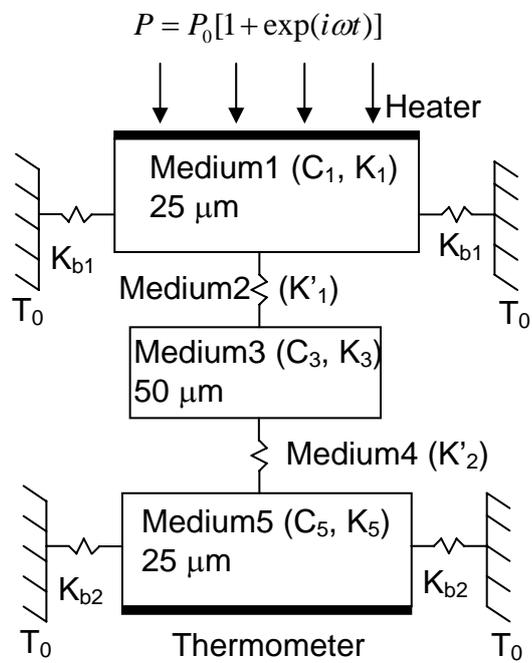

J.-L. Garden, Thermochimica Acta Figure 2

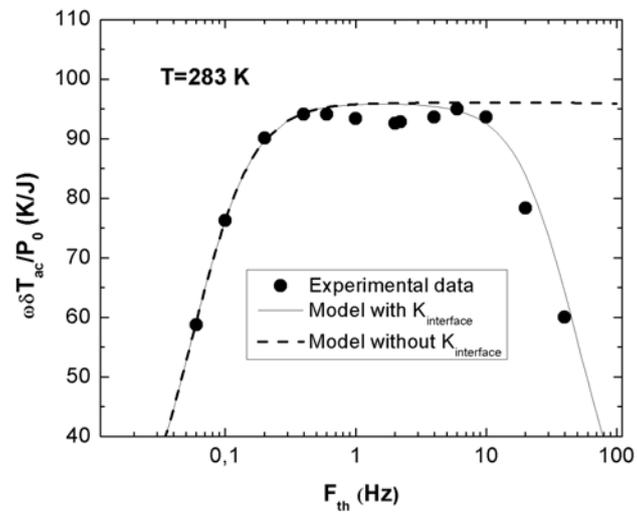

J.-L. Garden, Thermochimica Acta Figure 3

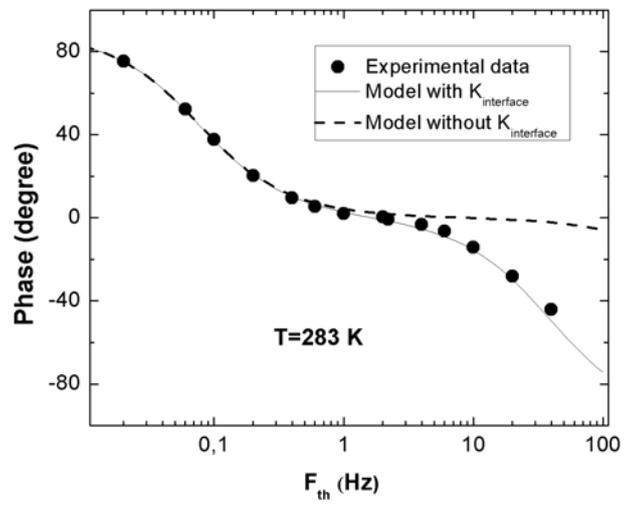

J.-L. Garden, Thermochimica Acta Figure 4

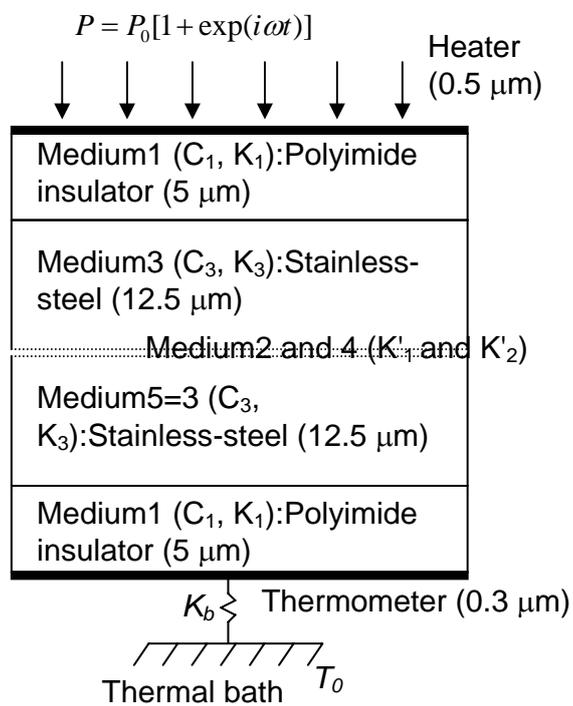

J.-L. Garden, Thermochimica Acta Figure 5

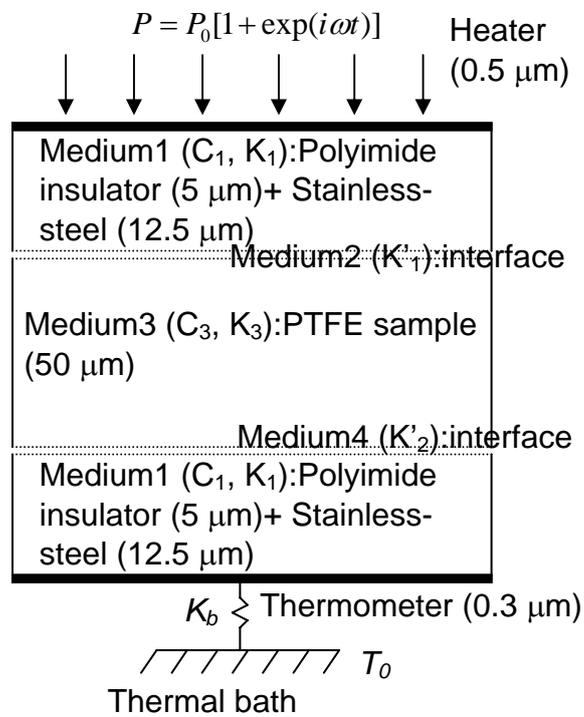

J.-L. Garden, Thermochimica Acta Figure 6

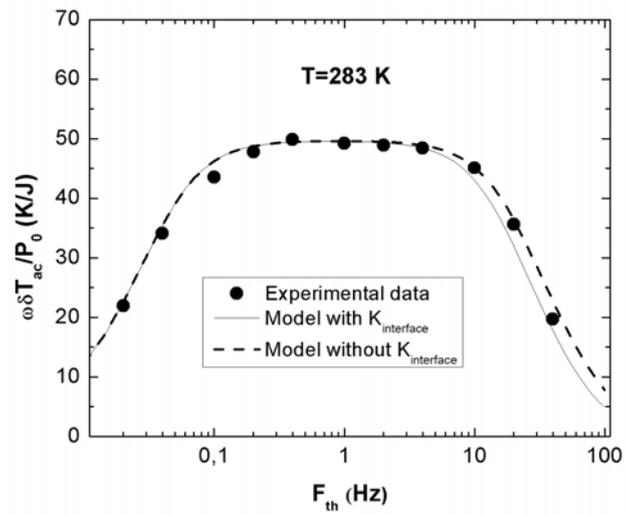

J.-L. Garden, Thermochimica Acta Figure 7

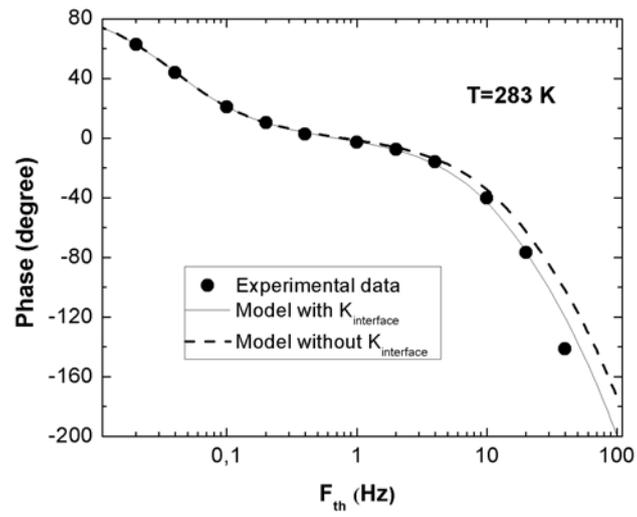

J.-L. Garden, Thermochimica Acta Figure 8

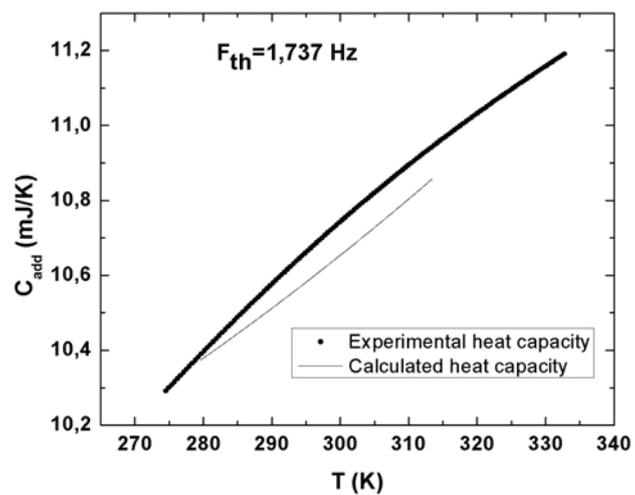

J.-L. Garden, Thermochimica Acta Figure 9

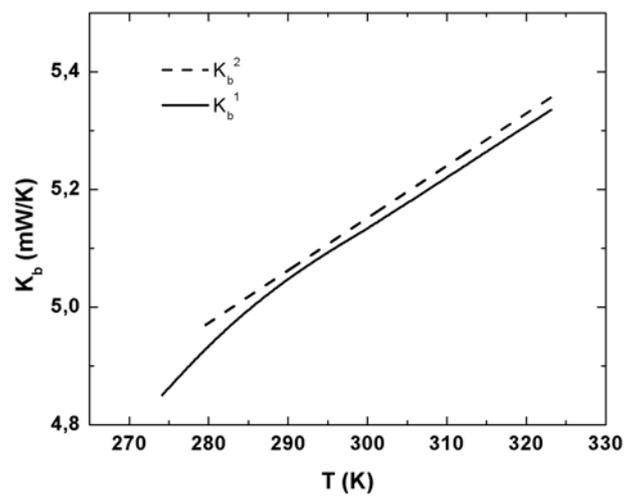

J.-L. Garden, Thermochimica Acta Figure 10

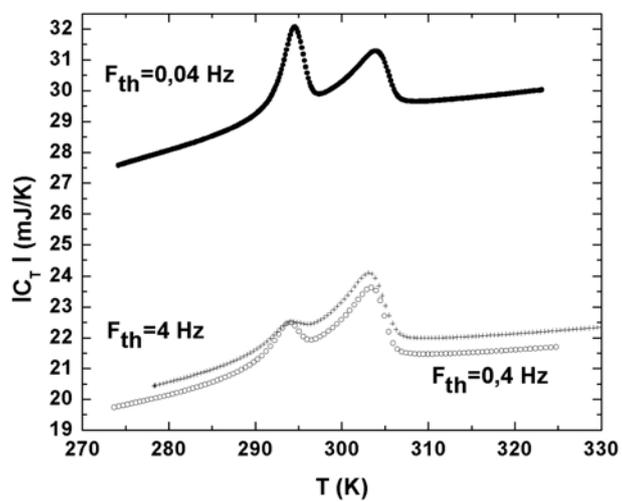

J.-L. Garden, Thermochimica Acta Figure 11

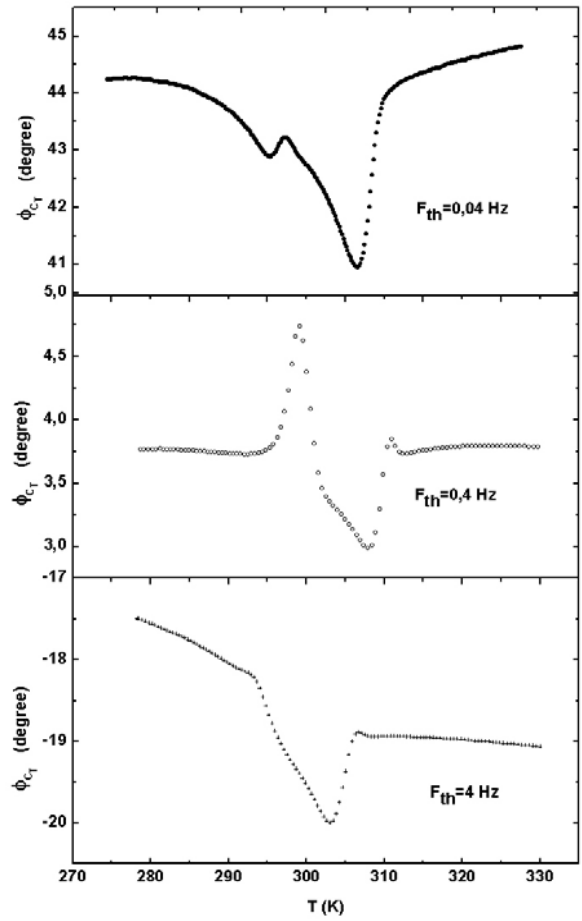

J.-L. Garden, Thermochimica Acta Figure 12

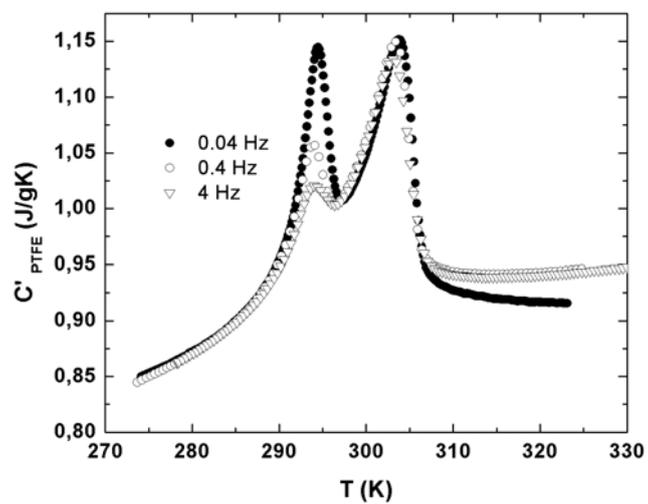

J.-L. Garden, Thermochimica Acta Figure 13

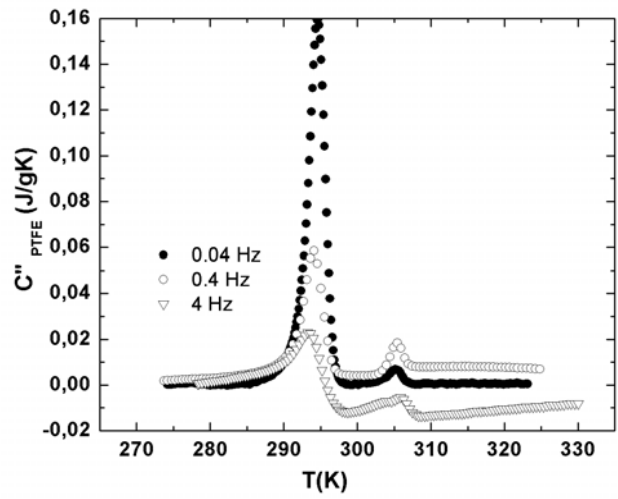

J.-L. Garden, Thermochimica Acta Figure 14

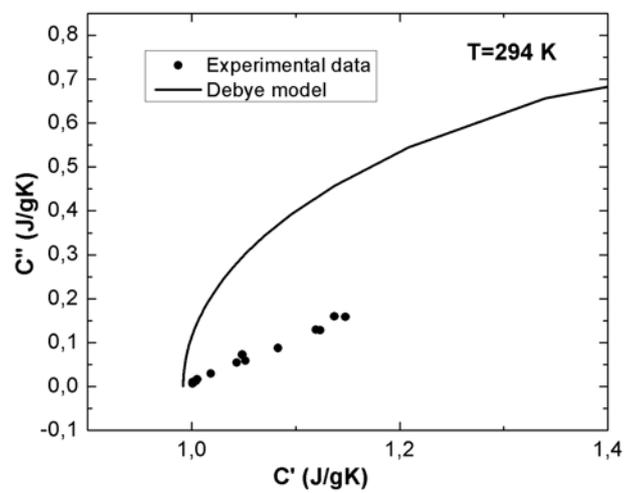

J.-L. Garden, Thermochimica Acta Figure 15

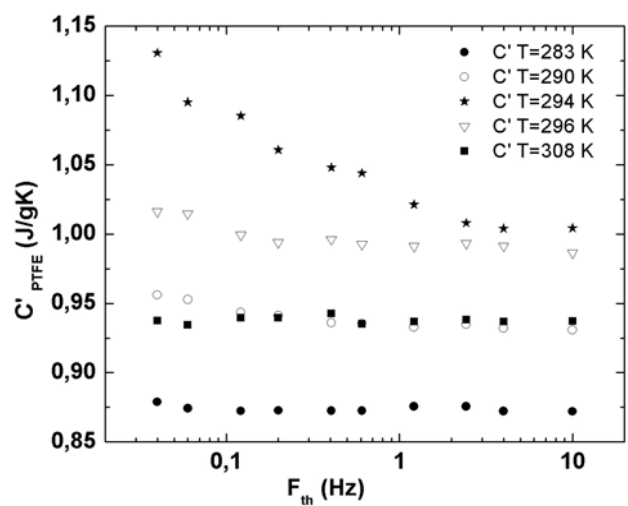

J.-L. Garden, Thermochimica Acta Figure 16

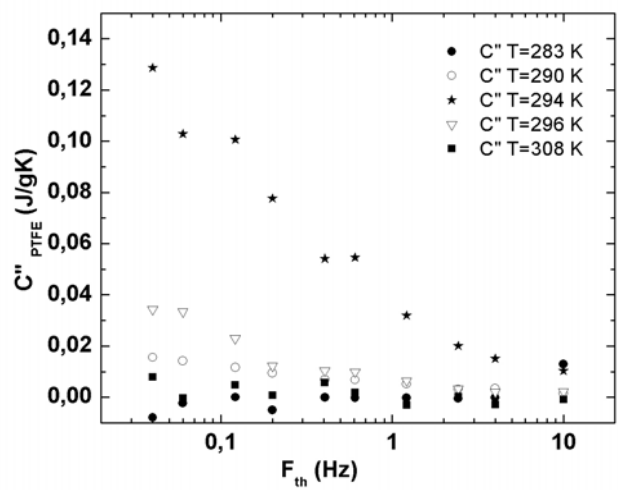

J.-L. Garden, Thermochimica Acta Figure 17

|  | Density(g.cm$^{-3}$) | $c_p$(J/g.K) T = 298 K | k(W/m.K) T = 298 K | $c_p$(J/g.K) T = 283 K | k(W/m.K) T = 283 K |
|---|---|---|---|---|---|
| Stainless-steel | 7.9 | 0.454 | 16 | 0.446 | 15.583 |
| Polyimide | 1.47 | 1.1268 | 0.2895 | 1.1020 | 0.28578 |
| PTFE |  |  | 0.25 |  | 0.2495 |

J.-L. Garden, Thermochimica Acta table1

|          | $C(J/K)$<br>T = 283 K | $K(W/K)$<br>T = 283 K | $\alpha/\sqrt{\omega}$<br>T = 283 K | $K'_1(W/K)$ | $K'_2(W/K)$ | $K_b(W/K)$ |
|----------|-----------------------|-----------------------|-------------------------------------|-------------|-------------|------------|
| Medium1  | $0.801\times10^{-3}$  | 5.716                 | $8.4173\times10^{-3}$               |             |             |            |
| Medium3  | $8.8\times10^{-3}$    | 62.3328               | $8.40\times10^{-3}$                 |             |             |            |
| Fit 1    |                       |                       |                                     | 1000        | 1000        | $5\times10^{-3}$ |
| Fit 2    |                       |                       |                                     | 0.17        | 1000        | $5\times10^{-3}$ |

J.-L. Garden, Thermochimica Acta table2

|  | $C(J/K)$ $T = 283$ K | $K(W/K)$ $T = 283$ K | $\alpha/\sqrt{\omega}$ $T = 283$ K | $K'_1(W/K)$ | $K'_2(W/K)$ | $K_b(W/K)$ |
|---|---|---|---|---|---|---|
| Medium1 | $5.226\times10^{-3}$ | 5.465 | $21.866\times10^{-3}$ |  |  |  |
| Medium3 | $9.598\times10^{-3}$ | 0.499 | $98.07\times10^{-3}$ |  |  |  |
| Fit 1 |  |  |  | 1000 | 1000 | $5\times10^{-3}$ |
| Fit 2 |  |  |  | 1.4 | 1000 | $5\times10^{-3}$ |

J.-L. Garden, Thermochimica Acta table3